\documentclass[sigconf,screen]{acmart} 

\AtBeginDocument{%
  }

\setcopyright{acmlicensed}
\copyrightyear{2024}
\acmYear{2024}
\acmDOI{10.1145/3640457.3688188}

\acmConference[RecSys '24]{18th ACM Conference on Recommender Systems}{October 14--18,
  2024}{Bari, Italy}
\acmISBN{978-1-4503-XXXX-X/18/06}

\copyrightyear{2024}
\acmYear{2024}
\setcopyright{rightsretained}
\acmConference[RecSys '24]{18th ACM Conference on Recommender Systems}{October 14--18, 2024}{Bari, Italy}
\acmBooktitle{18th ACM Conference on Recommender Systems (RecSys '24), October 14--18, 2024, Bari, Italy}\acmDOI{10.1145/3640457.3688188}
\acmISBN{979-8-4007-0505-2/24/10}

\newcommand{\diag}{\operatorname{diag}} 
\usepackage{multirow}
\begin{document}

\title[Enhancing Sequential Music Recommendation with Negative Feedback-informed Contrastive Learning]{Enhancing Sequential Music Recommendation with\\Negative Feedback-informed Contrastive Learning}

\author{Pavan Seshadri}
\orcid{0009-0008-7838-9614}
\email{pseshadri9@gatech.edu}
\affiliation{%
  \institution{Georgia Institute of Technology\\Music Informatics Group}
  \city{Atlanta}
  \state{Georgia}
  \country{USA}
}

\author{Shahrzad Shashaani}
\orcid{0000-0003-4344-2696}
\email{shahrzad.shashaani@tuwien.ac.at}
\affiliation{%
  \institution{TU Wien\\Faculty of Informatics}
  \city{Vienna}
  \country{Austria}
}

\author{Peter Knees}
\orcid{0000-0003-3906-1292}
\email{peter.knees@tuwien.ac.at}
\affiliation{%
  \institution{TU Wien\\Faculty of Informatics}
  \city{Vienna}
  \country{Austria}
}
\renewcommand{\shortauthors}{Seshadri et al.}

\begin{abstract}
Modern music streaming services are heavily based on recommendation engines to serve content to users. Sequential recommendation---continuously providing new items within a single session in a contextually coherent manner---has been an emerging topic in current literature. User feedback---a positive or negative response to the item presented---is used to drive content recommendations by learning user preferences. We extend this idea to session-based recommendation to provide context-coherent music recommendations by modelling negative user feedback, i.e., skips, in the loss function.

We propose a sequence-aware contrastive sub-task to structure item embeddings in session-based music recommendation, such that true next-positive items (ignoring skipped items) are structured closer in the session embedding space, while skipped tracks are structured farther away from all items in the session. 
This directly affects item rankings using a K-nearest-neighbors search for next-item recommendations, while also promoting the rank of the true next item. Experiments incorporating this task into SoTA methods for sequential item recommendation show consistent performance gains in terms of next-item hit rate, item ranking, and skip down-ranking on three music recommendation datasets, strongly benefiting from the increasing presence of user feedback. 
\end{abstract}




\maketitle

\section{Introduction}


The advent of music recommender systems (MRS)~\cite{schedl_etal:rshb:2022} has revolutionized the music industry, evidenced by their widespread integration into music streaming platforms. 
Sequential music recommendation, a core task within MRS, focuses on extending user sessions by suggesting the next track. Traditional methods rely on user history to learn preferences through techniques like collaborative filtering \cite{knees2013survey}, often facing the challenge of the cold start problem for new users or tracks \cite{schedl2015}. The model must guess preferences until the user and/or track has interacted with the system enough to learn a profile \cite{https://doi.org/10.48550/arxiv.1708.05031}. Sequential recommendation addresses this by learning session-level relationships, enabling more accurate predictions based on recent interactions. 

This study aims to leverage implicit and explicit within-context preference signals present in listening sessions to learn robust profiles for sequential recommendations. We investigate learning session-level information and the effects of incorporating user feedback through a contrastive learning task. We consider CNN-, RNN-, and transformer-based architectures, influenced by SoTA methods for sequential retail recommendation, and show a consistent improvement for all architectures through our extension.
To our best knowledge, learning from negative signals/user feedback has not been explored thoroughly for sequential music recommendation, most likely due to a lack of public data containing user feedback, especially time-aware feedback. 
In fact, many public music recommendation datasets were collected before the streaming boom, where logged listening histories would primarily be sourced from user selection, leading to a low source of negative signals ~\cite{schedl_etal:rshb:2022}.

To account for the change of listening behavior, in this work, we investigate three frequently used datasets with different characteristics across different periods, ranging from personal-collection-sourced listening profiles with few skips, to a large volume of streaming connected logs with increased skipping events, to a dedicated skip prediction dataset.
We show that an increasing amount of negative samples can be progressively exploited to learn effective session-level representations.




\section{Related Work}\label{sec:related_work}

Sequential recommendation systems can generally be divided into two types: \textit{session-aware} systems leverage session-level history from identifiable users, while \textit{session-based} systems ignore user-labels and aim to build user-agnostic representations using solely discrete sessions~\cite{10.1145/3109859.3109896}. 
In this study, we investigate a \textit{session-based} system that implicitly learns contextual information through anonymous listening sessions.

Several session-based approaches have been proposed for retail recommendation tasks. 
\emph{CASER} \cite{tang2018personalized} and \emph{NextItNet} \cite{10.1145/3289600.3290975} employ convolutional filters to learn effective  representations. 
\emph{GRU4Rec}~\cite{hidasi2015session} is a Recurrent Neural Network (RNN) recommender system to generate recommendations for short session-based data by modeling entire user sessions.
\emph{SLi-Rec}~\cite{yu2019adaptive} 
combines both long-term and short-term user modeling and introduces time-aware and content-aware controllers into the RNN structure.
Other session-based methods make use of attention mechanisms.
\emph{BERT4Rec}~\cite{https://doi.org/10.48550/arxiv.1904.06690} uses the bidirectional attention mechanism from BERT~\cite{https://doi.org/10.48550/arxiv.1810.04805} to learn a robust item embedding space for sequential recommendation. 
\emph{SASRec}~\cite{kang2018self} is a self-attention based sequential model to handle sparse datasets and capture longer-term semantics.

Sequential approaches have also been proposed for music recommendation tasks. 
These include \emph{CoSERNN}~\cite{10.1145/3383313.3412248}, which learns from contextual information such as device type and timestamps to learn embeddings, and Online Learning to Rank for Sequential Music Recommendation~\cite{10.1145/3298689.3347019}, which learns from music tags describing content information for online learning to rank.

Early work on incorporating negative user feedback for music recommendation has considered ad-hoc adjustments based on content and context similarity, e.g. \cite{elias_pampalk_2005_1414932,BOSTEELS20093342,yajie_hu_2011_1418301}.
In more recent work, performing a study close to our task, Wen et al.~\cite{10.1145/3298689.3347037} investigate leveraging implicit user feedback, immediately after clicking for video and music recommendations to improve WRMF~\cite{hu2008collaborative} and BPR~\cite{rendle2012bpr}.
Park and Lee~\cite{park2022exploiting} exploit item, document, or track negative feedback, and negative preference, for contrastive learning.

\looseness -1
In contrast to existing approaches, our method focuses on modeling context within sequences rather than relying completely on user taste or long-term preference profiles.
Since we formulate the core idea of our method as a learning objective via a loss term, it can be incorporated into a variety of existing recommendation algorithms.

\section{Method}\label{sec:method}
\subsection{Problem Statement}
\label{subection:probstat}

Given an item history sequence $S$, standard feedback-agnostic approaches to sequential recommendation aim to predict the next item $i$ at time $t+1$ for each $i \in S$. In other words, if $\textbf{F}(S)$ is a standard sequential recommender, operating on $S = \{i_0, i_1,..., i_t\}$, then $\textbf{F}(S) = \{i_1, i_2,..., i_{t+1}\}$. As this approach is feedback-agnostic, it is assumed that each item in $S$ is a valid recommendation.

Including user feedback, however, induces the possibility that certain items in $S$ may not be valid recommendations, and thus solely predicting the item at the next time step is not optimal. 
To combat this, we separate the items in $S$ with positive response $P$ and negative response $N$ into disjoint sets. 
We re-formulate the problem to instead desire the next-positive item for each $i_{t}$. 

If $j \in T_{p}$ denotes the set of time steps of positive items in $S$, then for each $i_{t}$, we aim to predict the next-positive item in $S$, $i_{m}$, such that $m = \min_{j}\{j \in T_{P} \mid j > t\}$.
The difference $m - t$ denotes the number of negative items between $i_{t}$ and its next positive item $i_{m}$. 
To learn $\textbf{F}(S)$, we model a probability distribution, \textbf{$p(t_{i+1} = i \mid S_{t}), \forall i \in I$}, which effectively provides a ranking of probabilities to select the desired next item. 
For each $i_{t} \in S$, we aim to raise the predicted probability of positive item $i_{m}$, while penalizing the probability of all negative items, $N = \{i_{t+n} \mid 0 < n < m - t\}$.


\subsection{Incorporating Negative Feedback}
For our core task of sequential recommendation, we target learning an inner product space, such that $\hat{\textbf{E}}\cdot\textbf{M}^T$ produces distances which can be mapped to the probability space described in Section \ref{subection:probstat}. Thus, these distances correspond to the ranking of each item during inference. To incorporate negative feedback, we propose a sequence-aware contrastive task used to affect these distances by regularizing the learned item embeddings, which in turn affects the ranking of positive items against negative items. 

While feedback-agnostic systems essentially model a similarity function where $\hat{\textbf{E}}(i_t) \sim \textbf{E}(i_{t+1})$, we aim to regularize the item embedding vectors to encourage our system to instead produce $\hat{\textbf{E}}(i_t) \sim \textbf{E}(i_{t+m})$, where $i_m$ is the next-positive example from $i_t$. 
To more robustly learn this embedding space, we also wish to treat all skipped tracks in a sequence as negative examples. 

To incorporate this into our model, we simply add an additional contrastive loss term to our learning objective. We employ InfoNCE noise-contrastive estimation \cite{DBLP:journals/corr/abs-1807-03748} to essentially model a classifier that is optimized to distinguish the next-positive item in the input sequence against all negative items in the sequence via a discriminative function, $f_{k}$. Utilizing cosine similarity as the discriminative function in the contrastive term directly augments the learned inner product space to promote the ranking of the true next-positive item against the rankings of all negative items. Formally, the loss term per item is shown below:
\begin{equation}
    \mathcal{L}_{NCE}=-\mathbb{E}_X\left[\log \frac{f_k\left(\textbf{i}_{m}, \hat{\textbf{i}}_{t+1}\right)}{f_k\left(\textbf{i}_{m}, \hat{\textbf{i}}_{t+1}\right) + \sum_{n_j \in N} f_k\left(n_{j}, \hat{\textbf{i}}_{t+1}\right)}\right]
\end{equation}
where $\textbf{i}_{m}$ is the embedding of the next-positive sample, and $\hat{\textbf{i}}_{t+1}$ is the predicted embedding of the next item. 

\section{Experiments}\label{sec:experiments}
\subsection{Models and Training}
We employ the following baseline models used in sequential item recommendation for our task: GRU4Rec~\cite{hidasi2015session}, CASER~\cite{tang2018personalized}, SASRec~\cite{kang2018self}, and BERT4Rec~\cite{https://doi.org/10.48550/arxiv.1904.06690}. 
We use these specifically for the variety in which sequential patterns are learned: by recurrent networks, convolutional filters unidirectional, and bidirectional attention, respectively. 
We aim to demonstrate that our loss function incorporating negative feedback can be generalized into a variety of architectures for sequential music recommendation. 

\subsubsection{Training Objective and Procedure}
We use equivalent implementations as laid out by the previous authors, except in sampling during training: empirically we find that using a sampled softmax with negative log likelihood improves training stability with a large vocabulary of items to recommend ($\sim$300-500K in this study). For each mini-batch for each session, we uniformly sample 1,000 unseen items and rank the target tracks alongside these. These 1,000 items are re-sampled each epoch, such that as training continues, the model continually learns to ``rank'' the target items with an increasing subset of the total tracks, as the number of unique tracks sampled for comparison increases. 

For our loss function, in line with other contrastive recommendation systems, we append the negative log-likelihood loss term for the sequential recommendation, with the contrastive loss term, shown below, where $\alpha$ and $\beta$ are constants:
$\mathcal{L} = \alpha \mathcal{L}_{NLL} + \beta \mathcal{L}_{NCE}$.


For all models except BERT4Rec, we train them to predict the next item at each time step. BERT4Rec uses bidirectional training with a masked language model, and a token is added at the end of each sequence to predict the next track based on the overall context. The final and second-to-last items in each session are used as testing and validation sets, respectively.


\subsubsection{Hyperparameters and Implementation}
We select a maximum sequence length of 20 for all datasets, breaking down longer sessions into multiple discrete sessions. 
For all models, the embedding and hidden dimensions are both set to 128. 
We refer to the base implementations as per the respective authors for the configuration of the GRU4Rec and CASER models. 
For both SASRec and BERT4Rec, we stack 2 self-attentive encoder blocks with 8 attention heads. We set the masking proportion for BERT4Rec to $p = 0.2$.  
All parameters are initialized via truncated normal sampling with $\mu = 0, \sigma = 1$ in range $[-0.02, 0.02]$.
We tune the optimal $\alpha, \beta \in [.1, .2, .5, .75, 1]$ using the validation set. For the MSSD, we select $\alpha = 1, \beta=0.5$, and for both LFM-1K and LFM-2B, we select $\alpha = 1, \beta=0.2$.
We use the ADAM optimizer \cite{DBLP:journals/corr/KingmaB14} with a learning rate of 0.005, selected after tuning through the validation set with $lr \in [0.0001, 0.0005, 0.001, 0.005]$.

\subsubsection{Non-sequential baseline competitors}
We also compare the models to non-sequential session-based baselines, namely Weighted Regularized Matrix Factorization (WRMF) \cite{hu2008collaborative} and Bayesian Personalized Ranking (BPR) \cite{rendle2012bpr}, and their negative feedback informed extensions proposed by Wen et al.~\cite{10.1145/3298689.3347037}. 
We evaluate both extensions: \emph{-BL} uses post-click feedback as corrections for preference labels; \emph{-NR} probabilistically
samples items across all types of feedback.

\subsection{Datasets}
For this study we use three real-world datasets that differ in terms of represented listening tasks, skipping activity, and years of listening activity contained, namely the Music Streaming Sessions Dataset (MSSD)~\cite{DBLP:conf/www/BrostMJ19} using data from Spotify, the LFM-2B dataset~\cite{schedl2022lfm}, and the LFM-1K dataset~\cite{Celma:Springer2010,oscar_celma_2022_6090214}, both using data from Last.fm.\footnote{Spotify: \url{https://open.spotify.com}, Last.fm: \url{https://last.fm}}

\emph{MSSD} contains $\sim$160 million Spotify user sessions predominantly from the year 2018. 
We sample $\sim$600K discrete sessions containing $\sim$9M item interactions with $\sim$600K total unique tracks, containing roughly 50\% explicitly annotated skipped tracks.

The two Last.fm datasets contain user listening histories with timestamp information, which we split into sessions wherever timestamp differences over 20 mins occur. 
We interpret a track to have been skipped if the next event starts within 30 secs after the track has been started. 
We remove any session with less than 5 events.
\emph{LFM-1K} contains 19 million discrete listening events from 2005 to 2009 (i.e., before streaming was the dominant way of listening) with only a 0.8\% skip rate, insufficient to learn from negative feedback. Thus, we sample sessions such that roughly half will contain a skip event. This provides a set of $\sim$80K sessions spanning $\sim$1.4M item interactions, containing $\sim$300K tracks with a skip rate of $\sim$11\%. 
\emph{LFM-2B} contains more than 2 billion listening events from 2005 until 2020. 
We specifically use the provided 2020 subset containing $\sim$7.6M listening events. After processing, the used subset contains $\sim$14\% skipped tracks.
We assume that this subset already has a stronger connection to streaming-based listening behavior than LFM-1K.
In all datasets, we discard all present user information and treat each session individually, i.e., 
as a new user.

\subsection{Evaluation}
For evaluating the core recommendation task of next predicting a desired item, we measure Hit Rate~$@\{1, 5, 10, 20\}$ and Mean Average Precision~$@ 10$ to emphasize the top recommendations. 
This common evaluation strategy alone has the limitation the next-positive-item is assumed to be the only true desired output of a recommender. 
Moreover, in the case of discrete sessions, the final (and penultimate, etc.) items in the sequence may not have a next-positive example recorded in the dataset.
Second, evaluation should reflect whether the penalization of negative items relative to positive items improves the learning of session-context overall to better predict unobserved items. 
We, therefore, evaluate an additional skip down-ranking task to quantify the reduction in the ranking of skipped tracks utilizing Mean Reciprocal Rank~$@ 10$.
A lowered MRR value for skips is considered an improvement and serves as a proxy for a better learned representation of context.

\section{Results and Discussion}\label{sec:results}

\begin{table*}
\centering
\begin{tabular}{llllllllllllllll}
\toprule
\multirow{2}{*}{\rotatebox[origin=c]{90}{\textbf{Data}}} & \textbf{Metric} &  \multicolumn{2}{l}{\textbf{SASRec}} & \multicolumn{2}{l}{\textbf{BERT4Rec}} & 
\multicolumn{2}{l}{\textbf{GRU4Rec}} & 
\multicolumn{2}{l}{\textbf{CASER}} &
\multicolumn{3}{l}{\textbf{WRMF}} & 
\multicolumn{3}{l}{\textbf{BPR}} \\ 
& & \emph{orig.} & \emph{ours} & \emph{orig.} & \emph{ours} & \emph{orig.} & \emph{ours} & \emph{orig.} & \emph{ours} & \hspace{1mm}\emph{orig.} & \hspace{-1mm}\emph{-BL} & \hspace{-1mm}\emph{-NR} & \hspace{1mm}\emph{orig.} & \hspace{-1mm}\emph{-BL} & \hspace{-1mm}\emph{-NR}\\
\midrule
\multirow{5}{*}{\rotatebox[origin=c]{90}{\textbf{MSSD}}} & \textbf{HR@1} & .377 & \textbf{.410} \textbf{(9\%)} & .204 & .355 \textbf{(74\%)} & .210 & .235 \textbf{(12\%)} & .223 & .251 \textbf{(13\%)} & \hspace{1mm}.203 & \hspace{-1mm}.207 & \hspace{-1mm}.211 & \hspace{1mm}.208 & \hspace{-1mm}.213 & \hspace{-1mm}.216 \\
 & \textbf{HR@5} & .615 & \textbf{.628} \textbf{(2\%)} & .450 & .608 \textbf{(35\%)} & .398 & .431 \textbf{(8\%)} & .412 & .455 \textbf{(10\%)} & \hspace{1mm}.400 & \hspace{-1mm}.406 & \hspace{-1mm}.411 & \hspace{1mm}.406 & \hspace{-1mm}.413 & \hspace{-1mm}.421 \\
 & \textbf{HR@10} & .696 & \textbf{.706} \textbf{(1\%)} & .553 & .693 \textbf{(25\%)} & .491 & .534 \textbf{(9\%)} & .518 & .530 \textbf{(2\%)} & \hspace{1mm}.488 & \hspace{-1mm}.498 & \hspace{-1mm}.505 & \hspace{1mm}.502 & \hspace{-1mm}.510 & \hspace{-1mm}.517 \\
 & \textbf{HR@20} & .767 & \textbf{.774} \textbf{(1\%)} & .648 & .767 \textbf{(18\%)} & .600 & .627 \textbf{(5\%)} & .616 & .649 \textbf{(5\%)} & \hspace{1mm}.597 & \hspace{-1mm}.603 & \hspace{-1mm}.608 & \hspace{1mm}.605 & \hspace{-1mm}.609 & \hspace{-1mm}.612 \\
 & \textbf{MAP@10} & .397 & \textbf{.417} \textbf{(5\%)} & .185 & .369 \textbf{(99\%)} & .176 & .193 \textbf{(10\%)} & .195 & .221 \textbf{(13\%)} & \hspace{1mm}.149 & \hspace{-1mm}.156 & \hspace{-1mm}.164 & \hspace{1mm}.160 & \hspace{-1mm}.168 & \hspace{-1mm}.173 \\
\midrule
\multirow{5}{*}{\rotatebox[origin=c]{90}{\textbf{LFM-2B}}} & \textbf{HR@1} & .190 & \textbf{.221} \textbf{(16\%)} & .101 & .117 \textbf{(16\%)} & .096 & .102 \textbf{(6\%)} & .102 & .112 \textbf{(10\%)} & \hspace{1mm}.097 & \hspace{-1mm}.098 & \hspace{-1mm}.102 & \hspace{1mm}.098 & \hspace{-1mm}.105 & \hspace{-1mm}.107 \\
 & \textbf{HR@5} & .371 & \textbf{.400} \textbf{(8\%)} & .227 & .248 \textbf{(9\%)} & .203 & .221 \textbf{(9\%)} & .197 & .208 \textbf{(6\%)} & \hspace{1mm}.138 & \hspace{-1mm}.142 & \hspace{-1mm}.147 & \hspace{1mm}.143 & \hspace{-1mm}.152 & \hspace{-1mm}.169 \\
 & \textbf{HR@10} & .452 & \textbf{.477} \textbf{(6\%)} & .292 & .320 \textbf{(10\%)} & .273 & .291 \textbf{(7\%)} & .281 & .302 \textbf{(7\%)} & \hspace{1mm}.269 & \hspace{-1mm}.273 & \hspace{-1mm}.279 & \hspace{1mm}.276 & \hspace{-1mm}.286 & \hspace{-1mm}.294 \\
 & \textbf{HR@20} & .532 & \textbf{.553} \textbf{(4\%)} & .366 & .394 \textbf{(8\%)} & .311 & .342 \textbf{(10\%)} & .326 & .354 \textbf{(9\%)} & \hspace{1mm}.305 & \hspace{-1mm}.311 & \hspace{-1mm}.316 & \hspace{1mm}.314 & \hspace{-1mm}.324 & \hspace{-1mm}.348 \\
 & \textbf{MAP@10} & .188 & \textbf{.219} \textbf{(16\%)} & .098 & .110 \textbf{(12\%)} & .078 & .085 \textbf{(9\%)} & .081 & .088 \textbf{(9\%)} & \hspace{1mm}.062 & \hspace{-1mm}.065 & \hspace{-1mm}.067 & \hspace{1mm}.066 & \hspace{-1mm}.072 & \hspace{-1mm}.078 \\
\midrule
\multirow{5}{*}{\rotatebox[origin=c]{90}{\textbf{LFM-1K}}} & \textbf{HR@1} & .152 & \textbf{.181} \textbf{(19\%)} & .069 & .086 \textbf{(25\%)} & .048 & .059 \textbf{(23\%)} & .052 & .071 \textbf{(37\%)} & \hspace{1mm}.042 & \hspace{-1mm}.044 & \hspace{-1mm}.047 & \hspace{1mm}.043 & \hspace{-1mm}.050 & \hspace{-1mm}.052 \\
 & \textbf{HR@5} & .301 & \textbf{.330} \textbf{(10\%)} & .207 & .230 \textbf{(11\%)} & .182 & .200 \textbf{(10\%)} & .188 & .198 \textbf{(5\%)} & \hspace{1mm}.139 & \hspace{-1mm}.146 & \hspace{-1mm}.177 & \hspace{1mm}.150 & \hspace{-1mm}.153 & \hspace{-1mm}.159 \\
 & \textbf{HR@10} & .392 & \textbf{.421} \textbf{(7\%)} & .299 & .320 \textbf{(7\%)} & .261 & .289 \textbf{(11\%)} & .269 & .293 \textbf{(9\%)} & \hspace{1mm}.270 & \hspace{-1mm}.285 & \hspace{-1mm}.294 & \hspace{1mm}.292 & \hspace{-1mm}.298 & \hspace{-1mm}.301 \\
 & \textbf{HR@20} & .478 & \textbf{.491} \textbf{(3\%)} & .413 & .433 \textbf{(5\%)} & .388 & .397 \textbf{(2\%)} & .390 & .404 \textbf{(4\%)} & \hspace{1mm}.346 & \hspace{-1mm}.368 & \hspace{-1mm}.375 & \hspace{1mm}.369 & \hspace{-1mm}.374 & \hspace{-1mm}.376 \\
 & \textbf{MAP@10} & .092 & \textbf{.107} \textbf{(16\%)} & .049 & .064 \textbf{(31\%)} & .034 & .042 \textbf{(24\%)} & .037 & .040 \textbf{(8\%)} & \hspace{1mm}.028 & \hspace{-1mm}.031 & \hspace{-1mm}.034 & \hspace{1mm}.032 & \hspace{-1mm}.036 & \hspace{-1mm}.038 \\
\bottomrule
\end{tabular}
\caption{\emph{Hit Ratio @ [1, 5, 10, 20]} and \emph{Mean Average Precision @ 10} for the sequential models (SASRec, BERT4Rec, GRU4Rec, CASER) and the non-sequential baselines (WRMF, BPR) on the three datasets. Non \emph{``orig.''} models incorporate negative feedback. They are compared to their \emph{``orig.''} baselines which do not model negative feedback. Numbers in parentheses show the relative increase in the percentage of the approach over the respective baseline; bold entries mark the better performing approach between the baseline and negative feedback-informed approach. The overall best performance is also highlighted in bold (i.e., SASRec-ours).}
\label{tab:results_all_hr_map}
\end{table*}

\subsection{Core-Recommendation Task}

Table~\ref{tab:results_all_hr_map} shows the results for all models, comparing the baselines with negative feedback informed extensions (ours for the sequential approaches, \cite{10.1145/3298689.3347037} for the non-sequential) on all three datasets.
We find that our feedback-informed loss term unanimously increases performance across datasets. 
Additionally, for the non-sequential baselines (WRMF, BPR) we find improvements consistent with those reported in ~\cite{10.1145/3298689.3347037}. 

We observe the greatest absolute performance across all models on the MSSD, followed by the LFM-2B and LFM-1K, respectively. With regards to sequential music recommendation, the specific type of listening behavior and context could possibly affect the model's ability to model the interaction sequences effectively: the MSSD (collected 2019) is fully streaming based and contains a high proportion of algorithmically driven content and performs best, while the LFM-1K (collected 2009) contains mostly user curation and performs worst. LFM-2B (collected 2020) contains a mix of both and performs in the middle. 
Performance gains appear to correlate with skip percentage in the dataset as expected: With more available negative feedback, the feedback-informed loss term can better regularize embeddings. 

\subsection{Skip down-ranking}
Table~\ref{tab:results_all_mrr} details the results of the skip down-ranking experiment. 
We observe a consistent decrease in scores (lower is better) after learning from negative feedback, with a notable exception in BERT4Rec on MSSD, where we observe an increase. 
Additionally, this dataset contains the highest proportion of skipped tracks in both the testing and training sets (roughly 50\% for both), and this particular combination (BERT4Rec on MSSD) displayed the highest performance increase in the core-recommendation performance (18-74\% increase on hit rate). 

\begin{table}
\centering
\begin{tabular}{llllll}
\toprule
& &  \multicolumn{2}{l}{\textbf{SASRec}} & \multicolumn{2}{l}{\textbf{BERT4Rec}} \\ 
\textbf{Metric} & & \emph{orig.} & \emph{ours} & \emph{orig.} & \emph{ours} \\ 
\midrule
\textbf{MRR@10} & \textbf{MSSD} & .960 & .950 \textbf{(-1\%)} & .840 & .950 \textbf{(13\%)} \\ 
\textbf{MRR@10} & \textbf{LFM-2B} & .969 & .911 \textbf{(-6\%)} & .953 & .950 \textbf{(-0.3\%)} \\ 
\textbf{MRR@10} & \textbf{LFM-1K} & .731 & .731 \textbf{(0\%)} & .540 & .460 \textbf{(-15\%)} \\ 
\bottomrule
\end{tabular}
\caption{\emph{Mean Reciprocal Rank @ 10} on skip targets for the best sequential models (SASRec, BERT4Rec) on the three datasets. Lower values are better.}
\label{tab:results_all_mrr}
\end{table}

\subsection{Model Comparison}
Through our picked baselines, we observe multiple methods of learning from sequential interactions. SASRec and BERT4Rec both employ self-attentive architectures in uni- and bidirectional representations, resp. GRU4Rec uses recurrent networks, while CASER uses convolutional filters. We generally find superior performance with the self-attentive architectures, with SASRec consistently performing best across all datasets. For all datasets, we observe similar relative improvements in learning from negative feedback across all sequential models, except for BERT4Rec on the MSSD. 

While we observe the largest improvements in terms of HR and MAP, as per the results of the skip down-ranking, we also observe that skipped testing targets raise in rank. 
There are a couple of hypotheses that could support this: first, BERT4Rec is the only bidirectional approach of the 4 frontend sequential models. As we observe higher absolute performance consistently using SASRec (a similar architecture, differing in training objective for causal inference), the ``forward'' direction may be simply more impactful than the ``backward'' direction for sequential music recommendation. As our feedback-informed loss function explicitly models the forward direction, it could simply be enforcing this while training, leading to greater performance. Additionally, as the MSSD contains a high amount of algorithm driven content with user feedback, this particular combination may be learning the underlying algorithm rather than user intent. This could lead to significantly greater performance in predicting the end tracks, but also lead to making the same mistakes as the underlying algorithm (i.e., recommending tracks that will be skipped). 


\section{Conclusions}\label{sec:conclusion}
We have presented a contrastive learning objective that incorporates sequential negative feedback (i.e., skips) for next-track music recommendation.
We showed a consistent positive impact across three real-world datasets encompassing different listening behaviors, and across four SoTA sequential recommendation architectures. 
We additionally identify the ability of our proposed feedback-aware systems on reliably down-ranking unseen skipped tracks within the testing set. 

Direct future work would include the incorporation of explicit feedback (e.g. thumbs up/down) and hard negative/positive samples based on historical co-occurences.
Beyond the identified utility of incorporating negative feedback, our work points to open challenges to be addressed in future work, foremost a better understanding of user intents and different streaming behaviors, cf.~\cite{10.1145/3459637.3482123}, leading to more considerate models, both of data and in recommenders, and connected to that more specific evaluation strategies beyond single retrieval metrics focusing only on one aspect of systems.

\begin{acks}

This research was funded in whole or in part by the Austrian Science Fund (FWF) (\href{https://doi.org/10.55776/P33526}{10.55776/P33526}). For open access purposes, the author has applied a CC BY public copyright license to any author-accepted manuscript version arising from this submission.
\end{acks}

\bibliographystyle{ACM-Reference-Format}
\bibliography{ref}

\end{document}